\documentclass[prl,notitlepage,showpacs,amsmath,amssymb,twocolumn,groupedaddress,longbibliography]{revtex4-1}
\usepackage[utf8]{inputenc}
\usepackage{latexsym}
\usepackage{graphicx}
\usepackage{color}
\usepackage{amsmath}

\AtBeginDocument{%
    \newwrite\bibnotes
    \def\bibnotesext{Notes.bib}
    \immediate\openout\bibnotes=\jobname\bibnotesext
    \immediate\write\bibnotes{@CONTROL{REVTEX41Control}}
    \immediate\write\bibnotes{@CONTROL{%
    apsrev41Control,author="08",editor="1",pages="1",title="0",year="1"}}
     \if@filesw
     \immediate\write\@auxout{\string\citation{apsrev41Control}}%
    \fi
}%

\begin{document}

\title{Phase Signature of Topological Transition in Josephson Junctions} 

\author{Matthieu C. Dartiailh$^{1}$}
\author{William Mayer$^{1}$}
\author{Joseph~Yuan$^{1}$}
\author{Kaushini~S.~Wickramasinghe$^{1}$}
\author{Alex Matos-Abiague$^{2}$}
\author{Igor \v{Z}uti\'c$^{3}$}
\author{Javad~Shabani$^{1}$}
\affiliation{$^{1}$Center for Quantum Phenomena, Department of Physics, New York University, NY 10003, USA}
\affiliation{$^{2}$Department of Physics \& Astronomy, Wayne State University, Detroit, MI 48201, USA}
\affiliation{$^{3}$Department of Physics, University at Buffalo, State University of New York, Buffalo, New York 14260, USA}

\begin{abstract}

    Topological superconductivity holds promise for fault-tolerant quantum computing. While planar Josephson junctions are attractive candidates to realize this exotic state, direct phase-measurements as the fingerprint of the topological transition are missing. By embedding two gate-tunable Al/InAs Josephson junctions in a loop geometry, we measure a $\pi$-jump in the junction phase with increasing in-plane magnetic field, ${\bf B}_\|$. This jump is accompanied by a minimum of the critical current, indicating a closing and reopening of the superconducting gap, strongly anisotropic in ${\bf B}_\|$. Our theory confirms that these signatures of a topological transition are compatible with the emergence of Majorana states.

\end{abstract}

\maketitle 

Majorana bound states (MBS), which are their own antiparticles, are predicted to emerge as zero-energy modes localized at the boundary between a topological superconductor and a topologically-trivial region ~\cite{kitaev_unpaired_2001}. MBS can nonlocally store quantum information and their non-Abelian exchange statistics allows for the implementation of quantum gates through braiding operations ~\cite{nayak_non-abelian_2008}. This makes them ideal candidates for robust qubits in fault-tolerant topological quantum computing~\cite{aasen_milestones_2016}. Rather than seeking elusive spinless p-wave superconductors required for MBS, a common approach is to use conventional s-wave superconductors to proximity-modify semiconductor heterostructures with suitable symmetries~\cite{fu_superconducting_2008}.

Early MBS proposals were focused on one-dimensional (1D) systems such as proximitized nanowires and atomic chains ~\cite{lutchyn_majorana_2010,oreg_helical_2010,mourik_signatures_2012,rokhinson_fractional_2012,nadj-perge_observation_2014}, where the observation of a quantized zero-bias conductance peak (ZBCP)~\cite{sengupta_midgap_2001} provided the support for MBS. However, the inherent difficulties in the technological implementation of the required networks and the intrinsic instabilities of their 1D elements have motivated the search for versatile 2D platforms using more conventional devices such as Josephson junctions (JJs) and spin valves~\cite{shabani_two-dimensional_2016,suominen_anomalous_2017,ren_topological_2019,fornieri_evidence_2019,fatin_wireless_2016,matos-abiague_tunable_2017,ke_ballistic_2019,hart_controlled_2017}. Recent experiments~\cite{ren_topological_2019,fornieri_evidence_2019} suggest that planar JJs are particularly promising because they support topological superconductivity over a large parameter range. The change between trivial and topological superconductivity, probed by ZBCP, is realized by applying in-plane magnetic field, ${\bf B_\|}$, and biasing the superconducting phase between 0 and $\pi$. This is achieved by embedding the JJ in a loop~\cite{ren_topological_2019} or by using two strongly-asymmetric JJs~\cite{fornieri_evidence_2019} in a superconducting quantum interference device (SQUID).

Since ZBCP could arise even without topological superconductivity, it is crucial to identify alternative signatures. A striking example is the closing and reopening of the superconducting gap with an increasing $B_\|$ that is simultaneously accompanied by a phase jump~\cite{fu_superconducting_2008,lutchyn_majorana_2010,oreg_helical_2010,hell_two-dimensional_2017,pientka_topological_2017,cayao_majorana_2017}. Direct phase measurements have proven to be a powerful probe to elucidate unconventional superconductivity~\cite{tsuei_pairing_2000}. Trivial Zeeman/exchange 0-$\pi$ transition can happen without spin-orbit coupling (SOC) ~\cite{yokoyama_anomalous_2014,kontos_2002_prl}, but is then expected to be independent of the ${\bf B_\|}$-direction, in contrast with the case of topological superconductivity. It was also proposed that a Zeeman-driven Fulde–Ferrell–Larkin–Ovchinnikov (FFLO)-like mechanism could lead to a minimum in the critical current, $I_c$, through a spatial variation of the superconducting order parameter~\cite{hart_controlled_2017}. This mechanism can be characterized by the required $B_\|$ to reach the first minimum,
\begin{equation}\label{bfflo_def}
    B_\mathrm{FFLO} = \frac{\pi}{2}\frac{\hbar\,v_F}{g\,\mu_B\,L},
\end{equation}
where $\hbar$ is the Dirac's constant, $v_F$ the Fermi velocity, $g$ the effective $g$-factor, $\mu_B$ the Bohr
magneton, and $L$ is the distance between the superconducting contacts.

In Fig.~1(a) we plot $B_\mathrm{FFLO}$ as a function of the distance between the contacts for JJs on HgTe~\cite{hart_controlled_2017}, InSb~\cite{ke_ballistic_2019}, Bi nanowires~\cite{murani_ballistic_2017} and InAs for a range of $v_F$ noted in the caption. This expression is in a good agreement with the published measurements (see also section 3 of the Supplemental Material~\cite{supplementary-materials}). In this work, we consider a device fabricated on InAs quantum well with the $g$-factor of 10 and the effective mass of $m^*\approx 0.03$ electron mass~\cite{shabani_two-dimensional_2016}. 
For $L=100\,$nm, using $|g| = 10$ and $n = 7\times 10^{11}\,\textrm{cm}^{-2}$,  $v_F =  \hbar\,\sqrt{2\,\pi\,n}/m^*\approx8\times10^5\,$m/s, we find from Eq.~(\ref{bfflo_def}) an estimate for the trivial $0-\pi$ transition at $B_\mathrm{FFLO} \approx 14.4$ T, still much higher than any $B_\|$ used in our study. 

\begin{figure}[htbp!]
\centering
\includegraphics[width=0.46\textwidth]{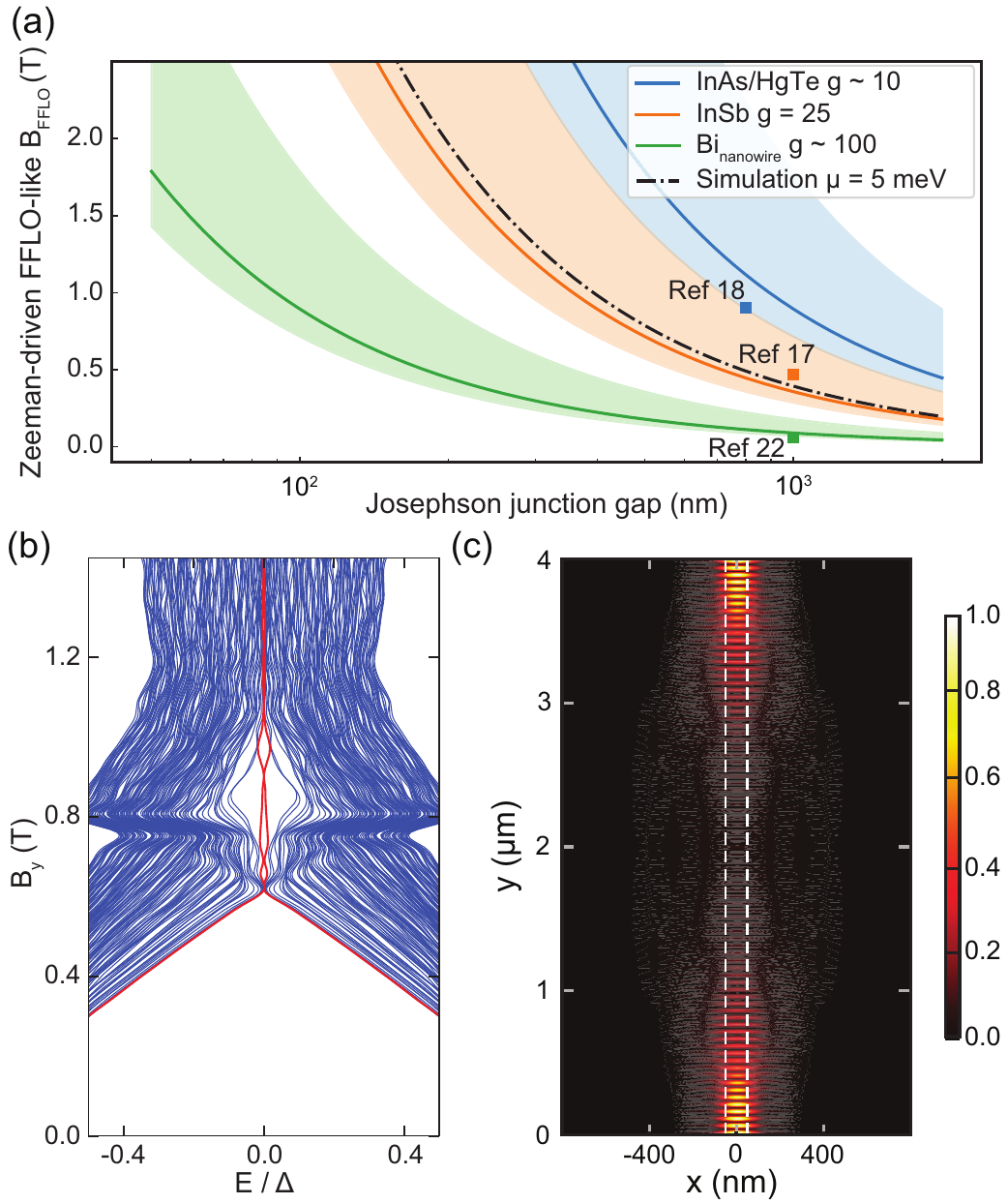}
\caption{\label{figure1}
(a) Predicted field at which a Zeeman-driven FFLO-like mechanism~\cite{hart_controlled_2017} leads to a minimum in $I_c$ as a function of the separation between the superconducting contacts. Solid line: $v_F=5\,\times 10^5$ m/s, colored region: from 4 $\times 10^5$ m/s to 1 $\times 10^6$ m/s. Dashed line: our simulation for which the FFLO-lilke mechanism would induce a transition at $\approx 4\,$T.
(b) 
The low-energy ground-state spectrum of a JJ with a semiconducting region at $\mu = 5.05$ meV. The energy gap closes and reopens indicating a topological phase transition and the emergence of Majorana bound states (MBS). Red lines: the evolution of finite-energy states into MBS inside the topological gap $\lesssim 23\,\mu$eV.
(c) Probability density of the MBS in the JJ for $B=0.85\,$T, normalized to its maximum value. It clearly indicates the formation of MBS localized at the end of the junction.  White lines: the edges of the normal region.
}
\end{figure}

With ${\bf B_\|}$ along the y-direction, tight-binding simulations presented in Figs.~1(b) and (c) predict a single junction, without a phase bias, will undergo a topological phase transition at $B_y > 0$. A hallmark of this transition is the closing and reopening of the superconducting gap~\cite{hell_two-dimensional_2017,pientka_topological_2017,cayao_majorana_2017,fu_superconducting_2008} manifested as a minimum in $I_c$ shown in Figs.~S8(a) and (d) 
of~\cite{supplementary-materials}. Above that closing, the system is in a topological phase dominated by chiral p-type superconductivity and host MBS localized at each end of the JJ as shown in Fig.~1(c). This is in contrast with a phase-biased junction~\cite{ren_topological_2019, fornieri_evidence_2019} which, at $\pi$-phase, can exhibit a topological phase at arbitrary low $B_y$.

Unlike the prior works~\cite{ren_topological_2019,fornieri_evidence_2019}, in our JJs the phase is not biased. Instead, it self-adjusts to minimize the free energy of the system in both the trivial and topological phases. We demonstrate how two gate-tunable and nearly-symmetric JJs, forming a SQUID provide a platform to realize the first direct phase-sensitive measurements of a topological transition. This is a complementary evidence for the topological nature of our observation of a minimum in $I_c$.  

The JJs based on epitaxial Al/InAs~\cite{shabani_two-dimensional_2016}  
are engineered to support high-interfacial transparency and high mobility resulting in JJ transparency $\tau\approx 0.9$, as estimated from the current-phase relation~\cite{mayer_gate_2019}, and robust proximity-induced superconductivity in InAs. Both junctions ($1$, $2$) of the SQUID are $W=4\,\mu$m wide and $L=100\,$nm long, while the area of the SQUID loop is $25\,\mu \textrm{m}^2$. The superconducting gap of the aluminium layer $\Delta_0 = 230 \pm 10 \,\mu$eV is estimated from measured critical temperature. With the coherence length $\xi_0=\hbar v_F/(\pi \Delta_0) \approx 770\,$nm and the mean free path, $l_e \approx 200\,$nm~\cite{mayer_gate_2019}, the JJs are in a quasi-ballistic short-junction limit, $l_e > L$ and $\xi_0 \gg L$~\cite{supplementary-materials,golubov_JJ_2004}. At high density ($n \approx 2\,\times\,10^{12}\,$cm$^{-2}$), based on previous study~\cite{wickramasinghe_transport_2018}, we expect the Rashba spin-orbit coupling (SOC) strength to reach $\alpha \approx 150\,$meV$\,$\AA, which corresponds to the spin-orbit energy,
$E_\mathrm{SO}=\alpha\, m^* v_F/\hbar \approx 5.3\,$meV.
We apply \textbf{B}$_\|$ along an arbitrary axis defined by angle $\theta$. At $\theta = 0$, the field is along y and  perpendicular to the current flowing through the JJ. We can also impose a phase difference between the two JJs by applying a small out-of-plane field. The versatility of our setup comes from the possibility to operate it either as a SQUID or as a single JJ by fully depleting the other JJ. 

\begin{figure}[htbp!]
\centering
\includegraphics[width=0.46\textwidth]{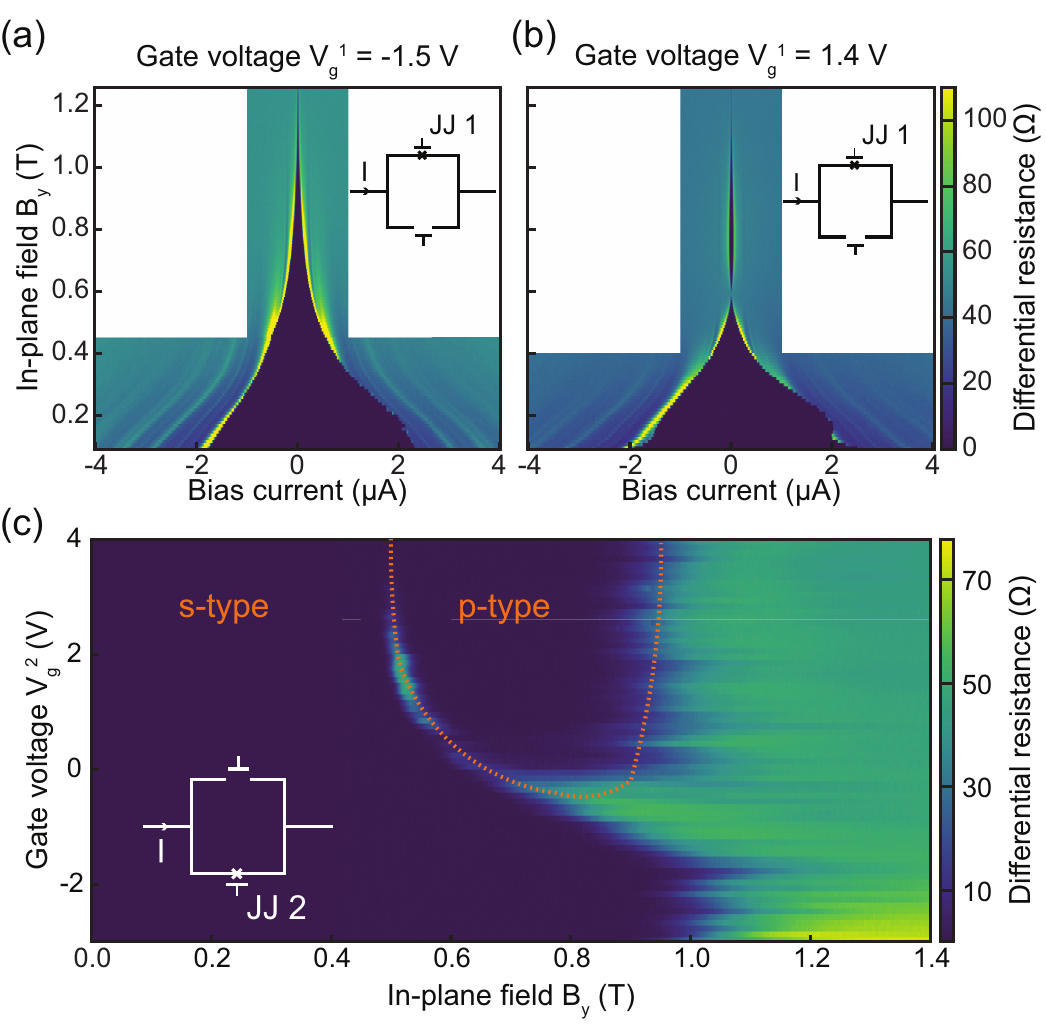}
\caption{\label{figure2}
Measurement of the differential resistance of JJ1 as function of an applied in-plane field along the y-axis at two different gate voltages (a) $V_g^1 =-1.5\,$V, (b) $V_g^1 = 1.4\,$V. In both cases, JJ2 is depleted ($V_g^2 = -7\,$V) and does not participate in the transport. At high gate (b), a minimum of the critical current is observed around $600\,$mT for JJ1.
(c), Zero-bias differential resistance of JJ2 as a function of the applied in-plane field and the gate voltage.
$V_g^1$ is set to $-7\,$V.}
\end{figure}

We next consider the $B_y$-dependence of the JJ1 critical current. In Fig.~2(a), at lower $V_g^1 = -1.5\,$V and thus at a lower density, we observe a trivial monotonic decrease of $I_c$ with $B_y$. Remarkably, at higher $V_g^1 = 1.4\,$V in Fig.~2(b), we see a striking difference with a non-monotonic behavior of $I_c$ and a minimum around $B_y = 600 \:\textrm{mT}$, in agreement with the tight-binding results from Fig.~S8(d)~\cite{supplementary-materials}.
Above that minimum, we measure $I_c \approx 20$ nA, consistent with the gap reopening and topological transition. After the minimum, the resistance vanishes for $I<I_c$. In contrast, after gap closing, only non-zero resistance was measured in Ref.~\cite{hart_controlled_2017}.

Minima in $I_c$ have been observed in Ref.~\onlinecite{fornieri_evidence_2019} due to orbital effects, which 
can become energetically unfavorable in the presence of strong proximity effect~\cite{setiawan_full_2019}. We expect that the proximity effect is stronger in our samples since the top barrier of our 2DEG is more transparent due to the lower band gap and electron mass of (In,Ga)As at 81\% In-content compared to the 75\% In-content used in \cite{fornieri_evidence_2019}. We discuss this point in greater details in \cite{supplementary-materials}.


Previous works~\cite{hell_two-dimensional_2017,pientka_topological_2017} have focused on the low-density and long-junction limit. In those works, the topological transition was found to happen at the Zeeman energy $E_\textrm{Z} = g\,\mu_B\,B_\|/2 \approx E_T/2$, where the Thouless energy is $E_T =(\pi/2)\hbar\,v_F/L$~\cite{pientka_topological_2017}. While this relation is equivalent to Eq.~(\ref{bfflo_def}), the Thouless energy in the ballistic  regime is not uniquely defined~\cite{ke_ballistic_2019,altland_nonstandard_1996}. These models do not apply to the short and high-$n$ junctions studied in this work for which $E_T\propto \sqrt{n}/L$ becomes larger ($\approx8.4\,$meV). The inadequacy of those models has also been pointed out in Ref.~\onlinecite{fornieri_evidence_2019}, which has motivated subsequent studies~\cite{setiawan_full_2019}, as well as in our own simulations~\cite{supplementary-materials}, confirming that the topological transition 
observed in our system is no longer determined by $E_Z = E_T/2$. However, we find that as $L$ increases, the transition field approaches the value predicted by Eq.~(\ref{bfflo_def})~\cite{supplementary-materials}. This is consistent with the derivation of Eq.~(\ref{bfflo_def}), where the effective $g$-factor was assumed to be finite only in the normal region. In our system we consider a finite $g$-factor over the whole junction. Therefore, the agreement with $E_Z = E_T/2$ is better when the normal region is long and the effects of finite $g$-factor in the proximitized InAs layer become less relevant.

In our device, both JJs show a nontrivial evolution of the superconducting gap and topological transition. Figure~2(c) shows the zero-bias differential resistance of JJ2 as a function $V_g^2$ and $B_y$. At the largest $V_g^2$, the transition occurs at $\approx 500$ mT and moves towards higher $B_y$, as $V_g^2$ is decreased. Below $V_g^2 = -1.5 \: \textrm{V}$, no evidence of any transition remains. The lower-$B_y$  transition in JJ2 compared to JJ1 can be attributed to small variation of junction properties, for example, lower supercurrent and the corresponding induced gap.

\onecolumngrid

\begin{figure}[b]
\centering
\includegraphics[width=0.92\textwidth]{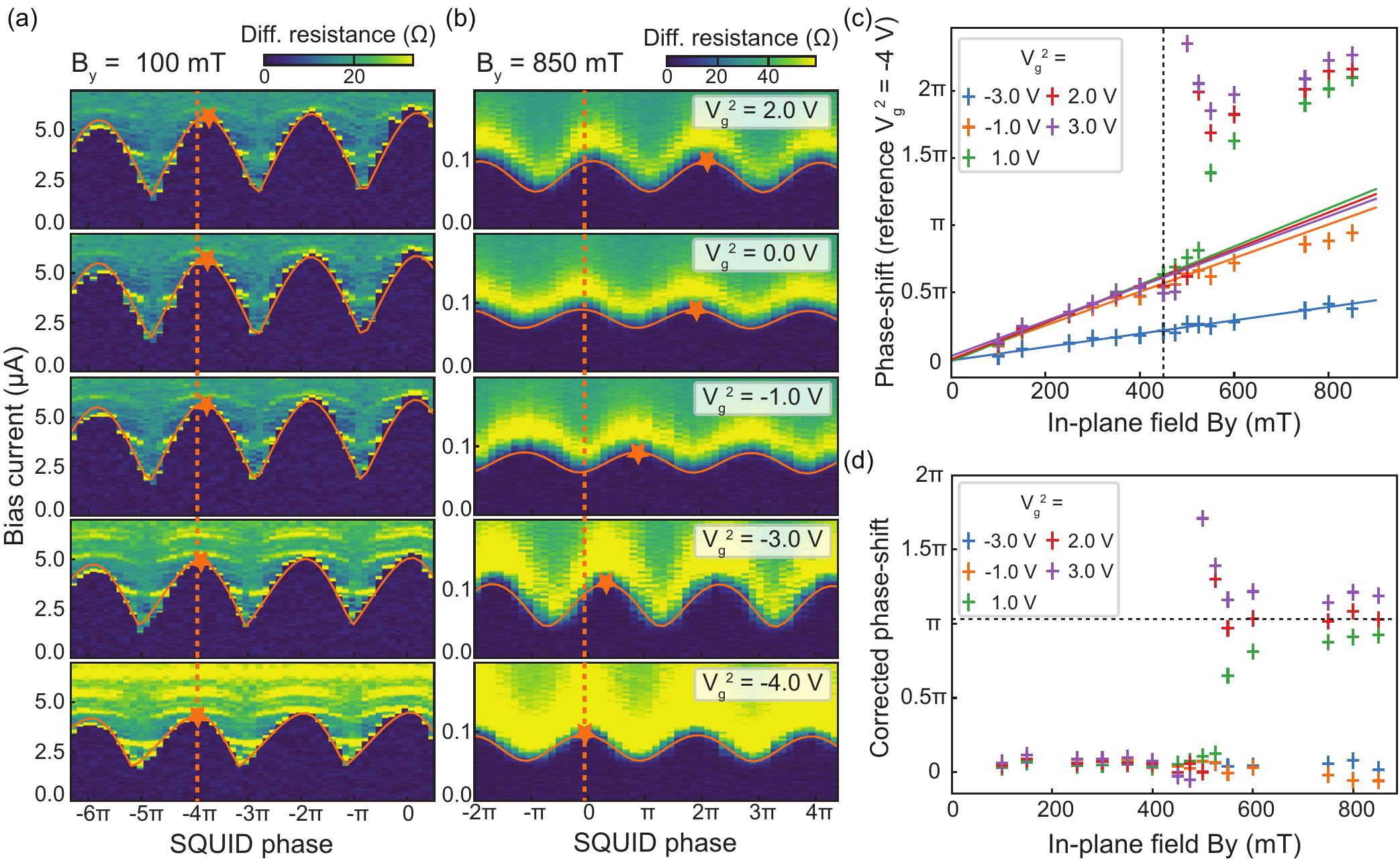}
\caption{\label{figure3} Phase signature of topological transition from SQUID interferometry. 
SQUID oscillations for $B_y=100\:\textrm{mT}$ (a) and $850\:\textrm{mT}$ (b) for different $V_g^2$ at $V_g^1=-2\:\textrm{V}$. Dashed lines: position of the maximum at $V_g^2=-4\:\textrm{V}$ used as a phase reference. Stars: position of the maximum of the oscillation. Solid lines: best fits to the SQUID oscillations used to extract the period and the phase-shift between different $V_g^2$.
(c) Phase difference between the SQUID oscillation at $V_g^2 = -4\:\textrm{V}$ and the oscillation at a different value as a function of $B_y$. Solid lines: linear fits of the data for $B_y \leq 450$ mT. (d) Phase shift from which the linear $B_y$-contribution has been subtracted to highlight the phase jump of about $\pi$ occurring for the three higher $V_g^2$ values.
}
\end{figure}

\twocolumngrid

While the observed non-monotonic dependence of $I_c$ with $B_y$ is consistent with a transition to topological superconductivity, phase-sensitive measurements with a SQUID could independently confirm this scenario. However, it is generally difficult to avoid arbitrary field offsets between measurements. Here, following the approach described in Ref.~\cite{mayer_gate_2019}, we use the gate tunability of our device to measure the phase offset between the oscillations observed at different gate voltage. We avoid the issue of the field offset by performing a single $B_z$ sweep and measuring $I_c$ at different gate voltage value for each value of the externally applied field. This procedure is valid, since changing the asymmetry between the $I_c$ of the two JJs does not affect the position of the maximum of the critical current in the SQUID~\cite{mayer_gate_2019}.

Using SQUID interferometry, we can identify a topological transition by setting JJ1 at $V_g^1=-2$ V as the reference junction. At this $V_g^1$, JJ1 does not show a topological transition at any $B_y$. The resulting SQUID oscillations in JJ2 reveal crucial differences between $B_y=100\:\textrm{mT}$ and $850\:\textrm{mT}$, shown respectively in Figs.~3(a) and (b), for various $V_g^2$. From Fig.~2(c), we expect that JJ2 would never reach the topological regime at $100\:\textrm{mT}$. Indeed, in Fig.~3(a), we only observe a small phase-shift which we attribute to the interplay between the $E_Z$ and SOC~\cite{yokoyama_anomalous_2014}. At higher $B_y$ in Fig.~3(b), there is a larger phase-shift between $V_g^2= -3$ V and -4 V than in Fig.~3(a), consistent with the expected linear increase in $B_y$~\cite{yokoyama_anomalous_2014}. However, comparing $V_g^2 = -1\:\textrm{V}$ and higher-gate values, a phase-shift of about $\pi$ occurs. 

Our tight-binding calculations, presented in Figs.~S8(a) and (d)~\cite{supplementary-materials}, reveal that a phase jump is a signature expected for the topological transition with the emergence of MBS. As shown, the ground-state phase (minimizing the energy in the absence of current) exhibits a phase jump when the topological transition occurs. Due to the broadening in the phase jump, the phase changes from 0 to a value close to $\pi$ not at a precise value of the magnetic field but over a finite range of $B_y$ [see Figs.~S8(a) and (d)]. The phase jump broadening may cause a shift between the magnetic field at which the system transits into the topological state and the field at which the minimum of $I_c$ occurs. Consequently, the topological transition occurs at a field smaller than or equal to the field corresponding to the $I_c$ minimum. The broadening of the phase jump observed in the experiment appears to be smaller that the one in the numerical simulations, suggesting the phase transition and the measured $I_c$ minimum occur at nearly the same $B_y$. 
In Fig.~3(c) we present the phase-shift between the reference scan performed at $V_g^2=-4\,$V and subsequent gate values. At low $B_y$, below the topological transition, we observe that the phase is linear in $B_y$, as indicated by the solid lines corresponding to linear fits to the values below 450$\,$mT. The increase in slope with  $V_g^2$ can be attributed to the increase of SOC~\cite{mayer_gate_2019}. For $V_g^2 = -3$ V and $-1$ V, the linear trend holds over all $B_y$. However, for $V_g^2=1\,$V, $2\,$V and $3\,$V, a jump can be observed around $550\,$mT, followed by another linear portion. To separate these effects, we subtract the linear component extracted from low-$B_y$ fits. The corrected phase in Fig.~3(d) reveals a phase jump with magnitude near $\pi$, around the $I_c$ minimum. This is a strong evidence for existence of topological phase transition consistent with theoretical calculations presented in \cite{supplementary-materials}.
The observed phase jump rules out orbital effects that could lead to a minimum in the $I_c$, as predicted in Ref.~\cite{pientka_topological_2017}. For a JJ with symmetric contact geometry, in this case identical, such effects would not lead to a phase jump and only the $I_c$ minimum would be observed.

\begin{figure}[htbp!]
\centering
\includegraphics[width=0.46\textwidth]{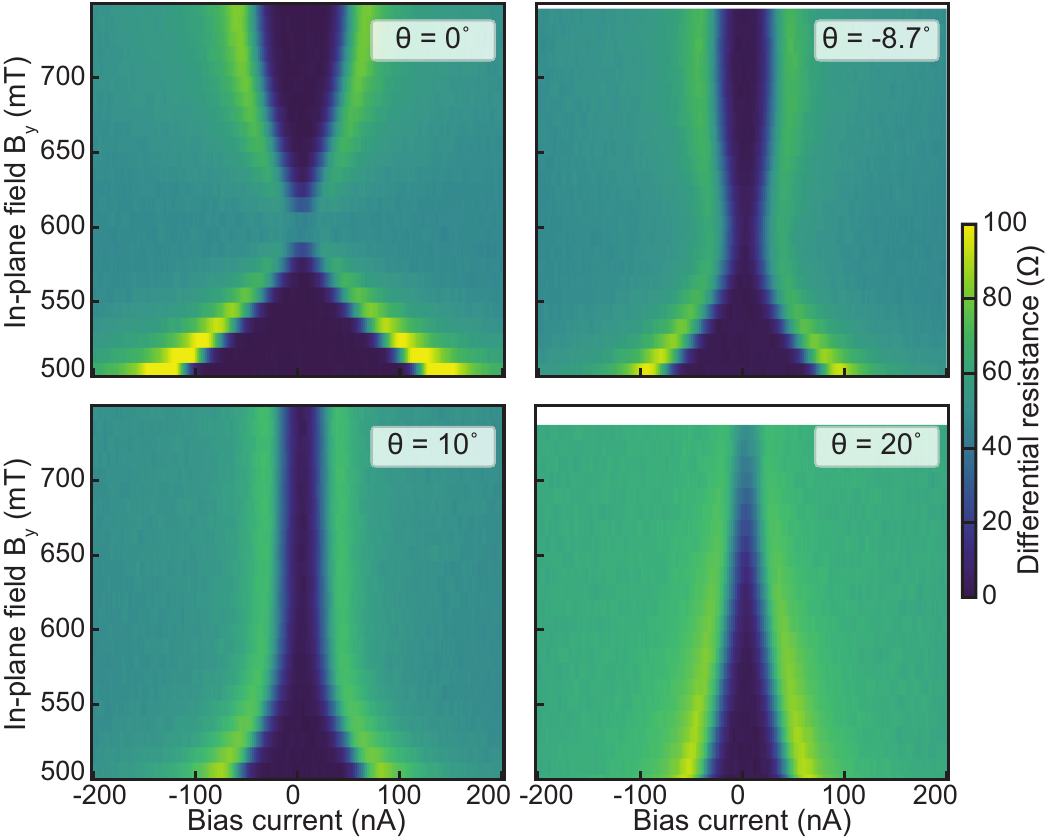}
\caption{\label{figure4}In-plane magnetic-field anisotropy of the gap closing. Differential resistance of JJ1 as function of the bias current and the 
y-component of ${\bf B}_\|$, applied at an angle $\theta$ with respect to the y-direction as depicted in Fig.~1(a).}
\end{figure}

 A distinct feature of the observed topological transition is its interplay of SOC and ${\bf B}_\|$. In our system, the topological regime is expected when ${\bf B}_\|$ is along the y-direction, i.e. $\theta=0$. Deviations from the y-direction results in a decrease of the topological gap for $\phi\lesssim\pi$~\cite{scharf_tuning_2019}. Therefore for large enough $\theta$, no topological transition nor associated minimum in $I_c$ is expected. We test this by probing the $I_c$ minimum  in a tilted ${\bf B}_\|$, away from $B_y$. In Fig.~4 we show $I_c$ zoom-ins of the JJ1 at $V_g^1=1.4\,$V. As $\theta$ is increased, the $I_c$ minimum increases but always occurs at the same value of $B_y$. If only the magnitude of $E_Z$ was relevant, one would expect to see the minima move to lower values of $B_y$. Similarly, SQUID data at $\theta = 10^{\circ}$~\cite{supplementary-materials}, display a reduced phase-shift, which may indicate a reduced topological gap. In contrast to smaller angles, at $\theta = 20^{\circ}$, $I_c$ decreases monotonically which suggests that s-wave order prevails and no transition is observed. This observation agrees with our calculations~\cite{supplementary-materials}. Unlike the insensitivity of trivial Zeeman $0-\pi$ transitions to the in-plane field rotation, our strong dependence on the ${\bf B_\|}$-direction is further evidence for a topological 0-$\pi$ transition. Furthermore, while for 2D InAs quantum wells only an out-of-plane anisotropy of the $g$-factor was measured~\cite{pakmehr_g-factor_2015}, we have examined that even the presence of a small in-plane $g$-factor anisotropy could not alone explain our results in Fig.~4.
 
 By embedding Al/InAs Josephson junctions in a nearly-symmetric SQUID loop, we measured two distinct signatures of a topological phase transition, a minimum in $I_c$ and the coincidental $\pi$-jump of the superconducting phase, both indicative of a closing and reopening of the superconducting gap. Our findings demonstrate the emergence of a topological phase. In addition to $B_y$, the top gate voltage is shown to be an efficient control knob for manipulating the topological phase transition. This offers a scalable platform for detection and manipulation of Majorana bounds states~\cite{zhou_phase_2020} and for development of complex circuits capable of fault-tolerant topological quantum computing.  The versatility of this two-dimensional geometry and SQUID sensing may also advance studies of MBS using magnetic textures for topological superconductivity~\cite{desjardins_synthetic_2019,yazdani_conjuring_2019} and support other exotic states that can be probed by phase-sensitive signatures~\cite{klinovaja_time-reversal_2014}.
\\


This work is supported by DARPA Topological Excitations in
Electronics (TEE) program.

\bibliography{lib}

\end{document}